\documentclass{PoS}
\pdfoutput=1

\usepackage{amsmath}
\usepackage{amssymb}
\usepackage{multirow}

\newcommand{\avg}[1]{\langle #1 \rangle}
\newcommand{\dd}{\mathrm{d}}

\title{Vector spectral functions and transport properties in quenched QCD}

\ShortTitle{Mesonic spectral functions and transport properties in quenched QCD}
\author{Heng-Tong Ding\\ 
	Key Laboratory of Quark \& Lepton Physics (MOE) and Institute of Particle Physics, Central China Normal University, Wuhan 430079, China \\
	E-mail: \email{hengtong.ding@mail.ccnu.edu.cn}}
\author{Olaf Kaczmarek\\ 
	Fakult\"at f\"ur Physik, Universit\"at Bielefeld, 
	33615 Bielefeld, Germany\\
	E-mail: \email{okacz@physik.uni-bielefeld.de}}
\author{\speaker{Florian Meyer}\\ 
	Fakult\"at f\"ur Physik, Universit\"at Bielefeld, 
	33615 Bielefeld, Germany\\
	E-mail: \email{fmeyer@physik.uni-bielefeld.de}}

\abstract{We present new results on the reconstruction of mesonic spectral
	functions for three temperatures $1.1T_c$, $1.2T_c$ and $1.4T_c$
	in quenched QCD. Making use of
	non-perturbatively improved clover Wilson valence quarks allows for a clean
	extrapolation of correlator data to the continuum limit. For the case of
	vanishing momentum the spectral function is obtained by fitting the data to
	a well motivated ansatz, using the full covariance matrix of the continuum
	extrapolated data in the fit. We found that vector correlation 
	function is almost temperature independent in the current 
	temperature window. The
	electrical conductivity of the hot medium, related to the origin of the
	vector spectral function at zero momentum, is computed from the resulting 
parameters at all three temperatures, leading to an estimate of 
$0.2C_{em}\lesssim \sigma/T\lesssim0.4C_{em}$. The dilepton rates resulting from 
the obtained spectral functions show no significant temperature dependence.}

\FullConference{The 32nd International Symposium on Lattice Field Theory,\\
23-28 June, 2014\\ Columbia University New York, NY}

\begin{document}

\section{Introduction} 
Ongoing Heavy Ion Collision experiments conducted at facilities like RHIC and
LHC provide new output about the nature of elementary particles and their
interactions. 
The spectral function in the vector channel at finite temperature provides
information on the thermal dilepton rates accessible in those 
experiments 
\cite{Rapp2009,Bernecker2011},
which we will attempt 
to extract from the fundamental theory of QCD in the following. Important
dynamical quantities can be extracted from the inherently non-perturbative 
regime of
small frequencies, which motivates the use of lattice data. 
With this we extend our former investigations \cite{Ding2011,Francis2012,Kacz2013}. 
A typically well accessible quantity on the lattice is the correlation function
in a given spectral channel. 
It inhibits dynamical properties of the QGP state when 
investigated at finite temperature \cite{Meyer2011,Hong2010}. 
As such, the light vector
correlator is related to the electrical conductivity $\sigma$ of the QGP,
the dilepton rate $\frac{\dd W}{\dd \omega\dd^3p}$ and the photon rate
$\frac{\dd R}{\dd^3p}$
as measured in heavy ion collision 
experiments, via its spectral function $\rho_V$ \cite{McLarren1984,Moore2007}.
While the spectral function relates to the correlator through an integral 
equation,
\begin{align}
	\label{eqn_integ_trans}
	G_H(\tau,\vec{p})= \int\limits_0^{\infty}
	\!\frac{\dd \omega}{2\pi}\rho_H(\omega,\vec{p},T) K(\omega,\tau,T) \quad
	\text{with} \quad
	K(\omega,\tau,T)=\frac{\cosh(\omega( \tau-\frac{1}{2T}))}{\sinh(\frac{\omega}{2T})},
\end{align}
the electrical conductivity is related to the spectral function via the 
Kubo formula,
\begin{align}
	\label{eqn_kubo}
	\frac{\sigma}{T}=\frac{C_{em}}{6}\lim\limits_{\omega\rightarrow 0} \frac{\rho_{ii}}{\omega}.
\end{align}
The two experimentally observable rates are in terms of the spectral 
function given by
\begin{align} 
	\frac{\dd W}{\dd \omega
	\dd^3p} \sim \frac{\rho_V(\omega,\vec{p},T)}
	{(\omega^2-\vec{p}^2)(e^{\omega/T}-1)},\quad
	\omega \frac{\dd R_{\gamma}} {\dd^3p}\sim\frac{\rho_V^T
	(\omega=|\vec{p}|,T)} {e^{\omega/T}-1}\;.
	\label{eqn_dilrate}
\end{align}
These relations imply that once the spectral function of the vector 
channel is extracted from QCD, important insights into non-perturbative phenomena
of heavy ion collisions and the QGP can be gained.
\newline

In order to determine the spectral function, however, (\ref{eqn_integ_trans}) 
has to be inverted,
which is often referred to as an "ill posed" problem \cite{Meyer2011}. 
The baseline of this reasoning is that the numerical (temporal) correlator 
data contains
$\mathcal{O}(10)$ points, while a decent resolution of the spectral function on
the other hand requires $\mathcal{O}(1000)$ points. Thus additional
information has to be 
provided, which we choose to be in the form of a phenomenologically
inspired ansatz
which is fitted to continuum extrapolated lattice QCD correlation functions.

\section{Lattice observables and continuum extrapolation} 
The renormalized isovector correlation function is constructed as 
\begin{align}
	J_{H}=Z_V \bar{\psi}(x)\gamma_{H}\psi(x) \quad \rightarrow
	\quad 
	G_{H}(\tau,\vec{x})=\avg{J_{H}(\tau, \vec{x})J^{\dagger}_{H}(0,\vec{0})},
	\label{eqn_correlator_xspace}
\end{align}
and projected to definite momentum $\vec{p}$:
\begin{align}
	\label{eqn_correlator_pspace}
	G_H(\tau,\vec{p}) =\sum_{\vec{x}} G_H(\tau,\vec{x})e^{i\vec{p}\vec{x}}.
\end{align}
In this study we constrain ourselves to the case $\vec{p}=0$. Splitting the 
correlation function (\ref{eqn_correlator_pspace}) into spatially and 
temporally polarized components, in Euclidean metric $G_V = G_{ii} + G_{00}$,
we form a ratio of correlation functions 
\begin{align}
	\label{eqn_ratio}
	R_{ii}=\frac{T^2}{\chi_q}\frac{G_{ii}(\tau T)} {G_{V}^{free,lat}(\tau T)}, \qquad \chi_q=-G_{00}/T,
\end{align}
where $G_{ii}$ is normalized by both the full free correlator on the 
lattice \cite{Karsch2003} and the 
quark number susceptibility $\chi_q$. The division by the latter rids us of 
the need to actually renormalize the spatial current correlator $G_{ii}$, while
the division by the former cancels its exponential falloff.
\newline

\begin{table}[h]
	\centering
	\begin{tabular}{|c||c|c|c|c|c|c|}
		\hline 
		& $\mathbf{N_{\tau}}$ & $\mathbf{N_{\sigma}}$ & $\mathbf{\beta}$ &
		$\mathbf{\kappa}$ & $\mathbf{1/a[GeV]}$ & $\mathbf{\#~conf.}$ \\ \hline \hline
		\multirow{3}{*}{$\mathbf{T=1.1T_c}$} &
		$32$ & $96$ & $7.192$ & $0.13440$ & $9.65$ & $314$ \\ 
				& $48$ & $144$ & $7.544$ & $0.13383$ & $13.21$ & $358$ \\ 
		  & $64$ & $192$ & $7.793$ & $0.13345$ & $19.30$ & $242$  \\ \hline
		\multirow{3}{*}{$\mathbf{T=1.2T_c}$} & 
		$28$ & $96$ & $7.192$ &$0.13440$ & $9.65$ & $232$ \\ 
				& $42$ & $144$ & $7.544$& $0.13383$ & $13.21$& $417$ \\ 
		  & $56$ & $192$ & $7.793$& $0.13345$ & $19.30$& $273$ \\ \hline
		\multirow{3}{*}{$\mathbf{T=1.4T_c}$} & 
		$24$ & $128$ & $7.192$ & $0.13440$ & $9.65$&$340$  \\
				& $32$ & $128$ & $7.457$ & $0.13390$ &$12.86$& $255$ \\ 
		  & $48$ & $128$ & $7.793$ & $0.13340$ &$19.30$& $456$ \\ \hline
	\end{tabular}
	\caption{Parameters of all lattices for all temperatures used in this study.}
	\label{tab_lattice_data}
\end{table}
Lattice calculations have been performed using a non-perturbatively improved 
Wilson-Clover action
without dynamical sea quarks at three different temperatures 
$T=1.1T_c,1.2T_c$ and $1.4T_c$ with 3 increasingly finer lattices each, see 
Tab.~\ref{tab_lattice_data}.
All valence quark masses 
are chosen to be small around $m_{\overline{MS}}(\mu=2GeV)\sim \mathcal{O}(10MeV)$. 
Note that for the two lowest temperatures the aspect ratio is fixed to
$N_s/N_t=3$ and $N_s/N_t=3.42$, respectively, ensuring a 
constant physical volume, while for the $T=1.4T_c$ lattice finite volume 
effects were verified to be small \cite{Ding2011}.
\newline

\begin{figure}[h]
	\centering
	\includegraphics[width=0.495\textwidth]{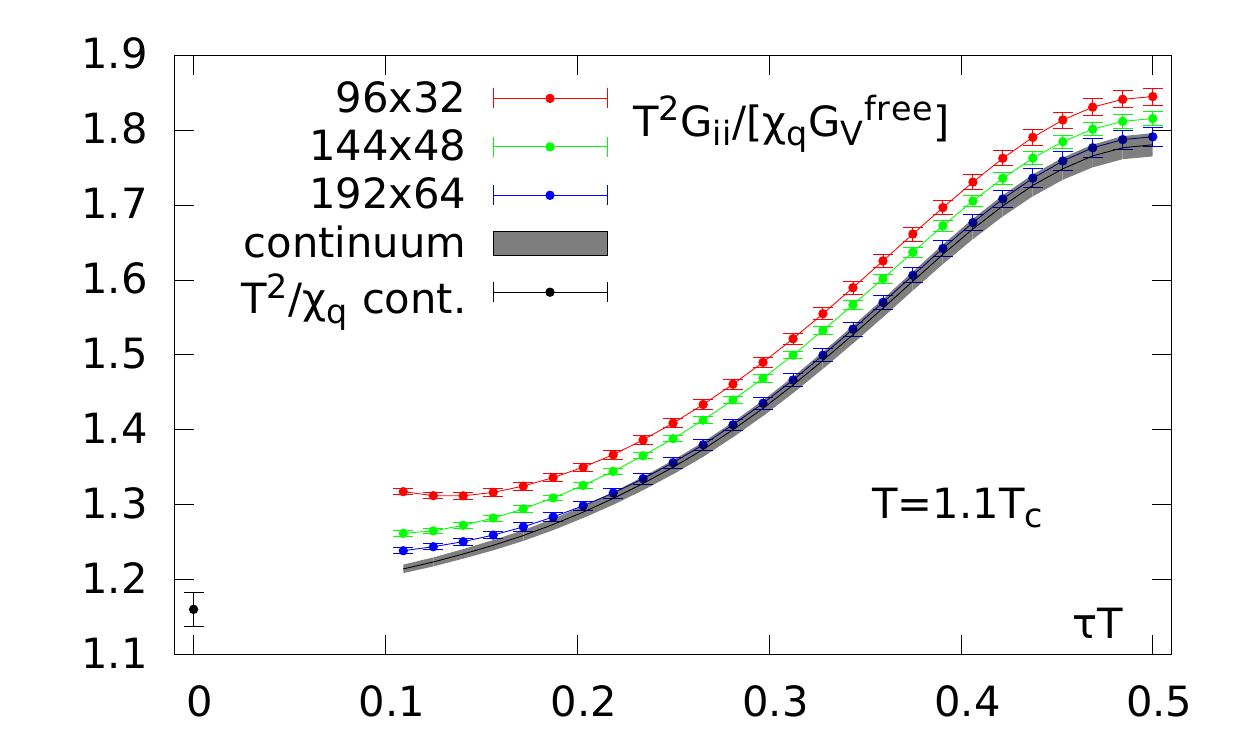}
	\includegraphics[width=0.495\textwidth]{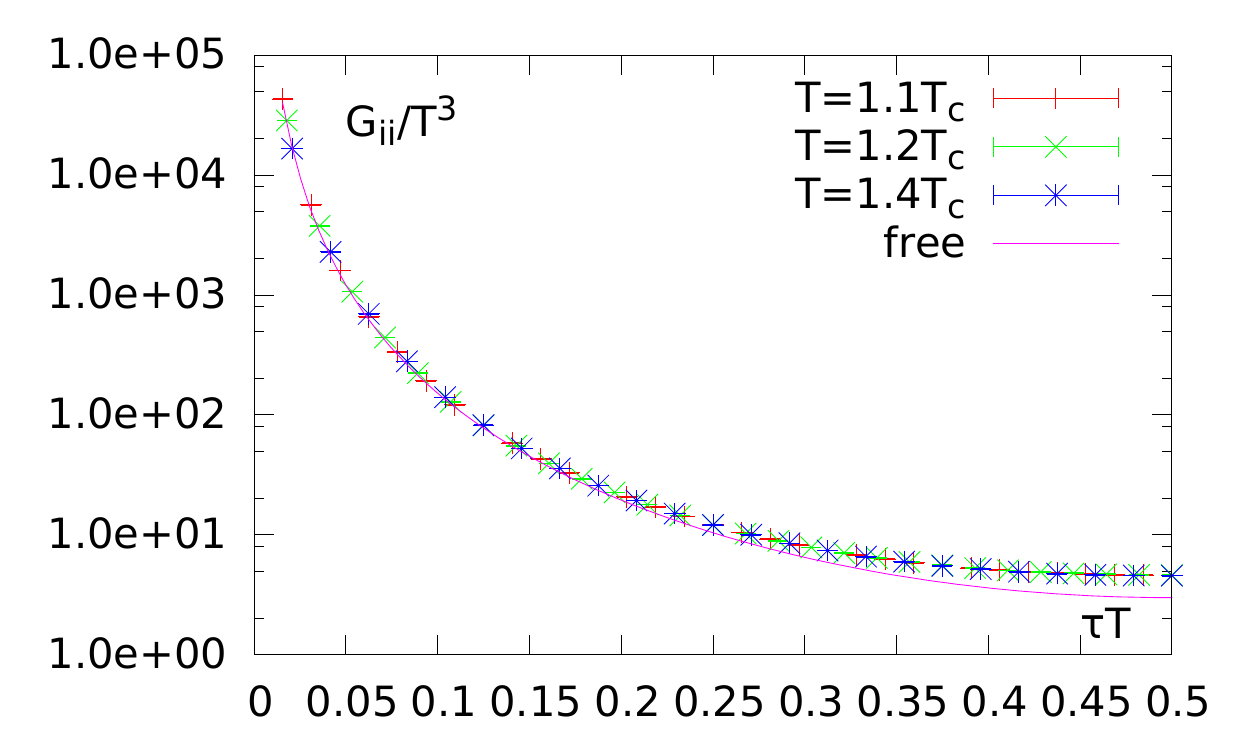}
	\caption{\textit{Left}: All three lattice correlators and the resulting 
		continuum extrapolated correlator for the $T=1.1T_c$ dataset. 
		Note that the finest lattice agrees with the continuum extrapolation 
		down to $\tau T\sim0.18$. The datapoint at
		$\tau T=0$ indicates the continuum extrapolated result for the
		inverted quark number suceptibility.
		\textit{Right}: Continuum extrapolated correlation functions for
		all three temperatures. Shown are the renormalized spatial 
	components. The solid line is the corresponding free correlation function.}
	\label{fig_cont_extr}
\end{figure}

For all three temperatures continuum extrapolations 
have been performed in $a^2$
for all $N_{\tau}/2$ original distances available on the finest lattice.
To achieve this, corresponding data 
points on the coarser lattices have been spline interpolated
along $\tau T$. The result is shown, for $T=1.1T_c$, in 
Fig.~\ref{fig_cont_extr} (\textit{left}).
The errors on the continuum extrapolated ratios obtained from a bootstrap analysis 
are slightly below the one percent level. 
The continuum extrapolated correlation functions $G_{ii}/T^3$ for each temperature
are shown in in Fig.~\ref{fig_cont_extr} (\textit{right}). The correlators 
overlap, thus we 
expect the same scaling with temperature in our resulting spectral functions,
already indicating that temperature effects in the dilepton rates and the 
electrical conductivities seem to be small.

\section{Fitting to the data}
In order to extract the vector spectral function via 
(\ref{eqn_integ_trans}) we employ an ansatz for its spatial part:
\begin{align}
	\rho_{ii}(\omega,T) &= \chi_q c_{\text{BW}} \frac{\omega\Gamma}{\omega^2+
	\label{eqn_ansatz}
(\Gamma/2)^2} + \frac{3}{2\pi}(1+k)\omega^2\tanh\left(\frac{\omega}
{4T}\right) \\
&\equiv \rho_{\text{BW}}(\omega,T) + (1+k)\rho_V^{\text{free}}(\omega,T).
\label{eqn_ansatz_short}
\end{align}
It consists of two parts: a Breit-Wigner peak, corresponding to the low $\omega$
region, and a modified version of the free spectral function. The 
modification parameter in the latter case fulfills $k=\alpha_s/\pi$ at 
leading order perturbation theory \cite{Ding2011}. This 
ansatz is inspired by the known relations for spectral functions in the 
non-interacting case, 
\begin{align}
	\rho_{ii}^{\text{free}}(\omega,T)=2\pi T^2\omega \delta(\omega) + \frac{3}{2\pi}
	&\omega^2\tanh(\frac{\omega}{4T}), \quad
	\rho_{00}^{\text{free}}(\omega,T) = 2\pi T^2 \omega \delta(\omega) \\
	\text{and}\quad
	\rho_{V}^{\text{free}}(\omega,T) &= \rho_{ii}^{\text{free}}(\omega,T)
	- \rho_{00}^{\text{free}}(\omega,T).
\end{align}
While the temporal correlator is constant due to charge conservation,
and thus the $\delta$-function in its spectral function is protected by 
symmetry, the corresponding $\delta$-function in the spatial part is expected
to be washed out upon the onset of interactions 
\cite{Hong2010,Moore2007,Aarts2002}.
This effect is hence modeled as a Breit-Wigner peak in our ansatz. 
\newline

An estimator for this spectral function is then obtained from relation 
(\ref{eqn_integ_trans})
by $\chi^2$-minimizing the ansatz on the r.h.s. with respect to the continuum 
extrapolated 
ratio data from eqn. (\ref{eqn_ratio}) on the l.h.s.
The fit itself is fully correlated
with the covariance matrix of the extrapolated continuum data estimated 
from the bootstrap samples.
From the entries of the covariance matrix it becomes apparent that 
there are covariances between data points used 
in the fit, which are comparable in size to the variances of the data at 
and around the 
midpoint, and hence non-negligible in the construction the $\chi^2$ function.
\newline 

However, the information about the small $\omega$ 
region resides in the large $\tau T$ region of the correlator 
\cite{Aarts2002II}, i.e.
around its midpoint. In order to extract more information from this region
we also extract and fit the first thermal moment of the correlator, 
see e.g. \cite{Ding2011} for a detailed discussion.

\section{Results}
In the following our procedure is shown using $T=1.1T_c$ as an 
example case.
The fits of the ansatz (\ref{eqn_ansatz}) to the continuum extrapolated data
show a very good convergence behaviour and 
yield as a result the three fit parameters $\Gamma$, $c_{BW}$, $k$
and their respective statistical fit errors, see the first column of 
Tab.~\ref{tab_fit_res_cut}. In Fig.~\ref{fig_cont_extr} (\textit{left})
one sees that the ratio on the finest lattice 
agrees with the continuum above $\tau T \simeq 0.18$, while cutoff effects are
visible for the coarser lattices also at larger distances. 
Although the continuum extrapolation seems to work also for smaller distances,
we are careful and include only those $\tau T$ in the fit where the finest
lattice agrees with the continuum extrapolation.
Generally, the smallest $\tau T$ to include in the fit for all temperatures 
lies in the interval $[0.18,0.20]$.
The value of $\chi^2/\text{dof}=1.24$ shows that the (rather simple) 
ansatz describes the data already well. 
The relative statistical fit errors of the parameters are roughly 
$30\%$ for $c_{BW}/\Gamma$ and $40\%$ for $\Gamma$. Note that the former 
has been calculated taking into account the correlation of the two 
parameters. 
\begin{figure}[h]
	\centering
	\includegraphics[width=0.495\textwidth]{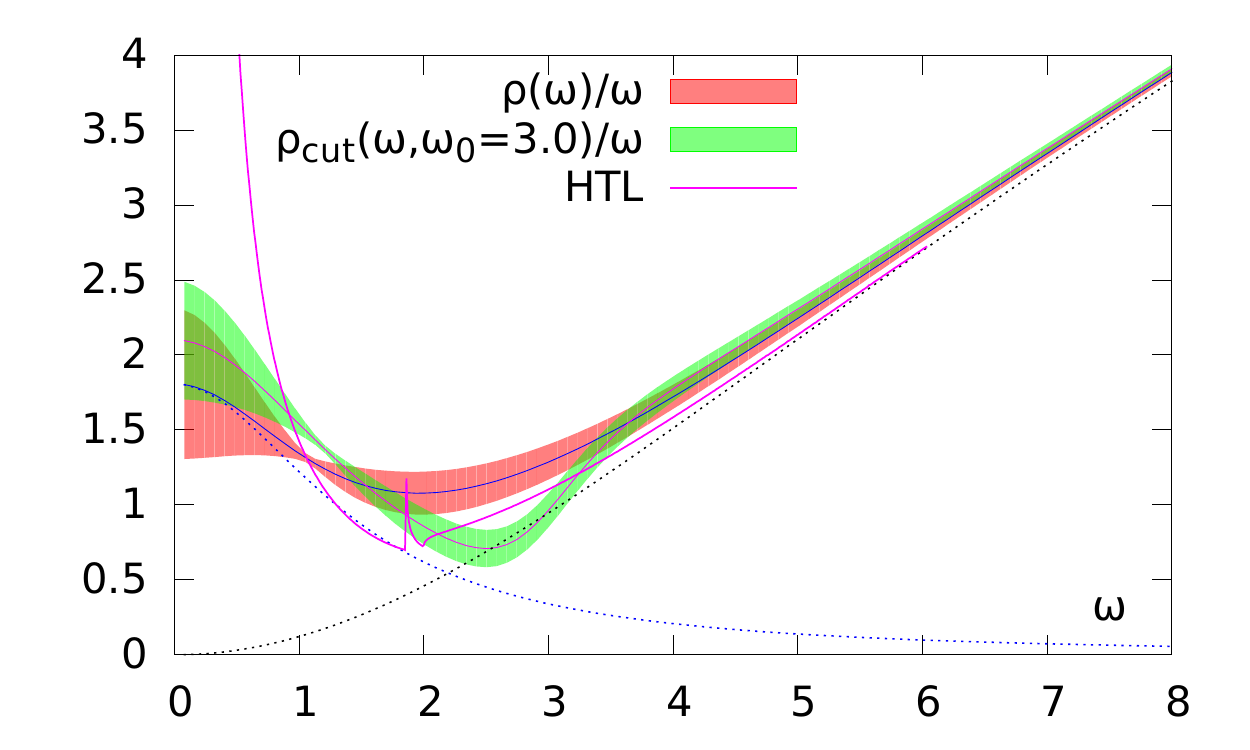}
	\includegraphics[width=0.495\textwidth]{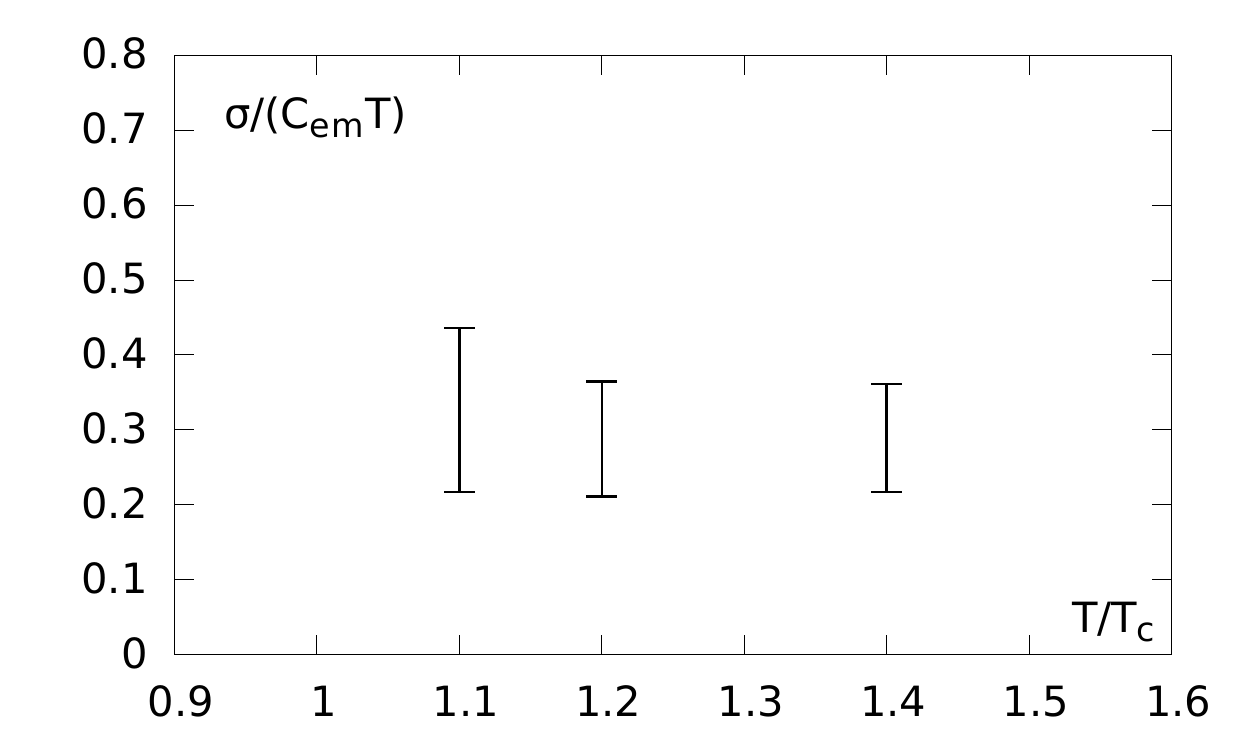}
	\caption{\textit{Left:} The spectral functions resulting from the fit for
		$T=1.1T_c$. The dotted lines are the Breit-Wigner and the free contributions
		seperately to guide the eye.
		\textit{Right:} The final results for the electrical conductivity for 
	all three temperatures. Their numerical values are listed in Tab.~3.}
	\label{fig_spf_plot}
\end{figure}
Using these parameters and their correlation matrix we construct
the resulting spectral function with its corresponding statistical 
errorband in Fig.~\ref{fig_spf_plot} (\textit{left}).
The electrical conductivity is then obtained from the origin of the 
spectral function via the Kubo relation (\ref{eqn_kubo}),
\begin{align}
	\frac{\sigma}{C_{em}T} =  \frac{2}{3T}\chi_q\frac{c_{BW}}
	{\Gamma}. 
\end{align}
\newline
\indent
In order to investigate a possible systematic uncertainty, we introduce
a low-frequency cutoff in the free part of the ansatz \cite{Ding2011},
\begin{align} 
	\label{eqn_cut_mod}
	\rho_V^{\text{free}}\rightarrow \rho_V^{\text{free}}\theta(\omega_0,\Delta_{\omega}) \quad \text{with} \quad 
	\theta(\omega_0,\Delta_{\omega}) = \left(1+\exp((\omega_0^2-\omega^2)/(\omega\Delta_{\omega}))
	\right)^{-1},
\end{align}
with $\theta(\omega_0,\Delta_{\omega})$ being a representation 
of the $\theta$-function for
$\Delta_{\omega}=0$, and smeared out for $\Delta_{\omega}\neq 0$.
Effectively, by varying $\omega_0$, we probe the sensitivity of our Ansatz
with respect to a continuous change in the low frequency region, i.e.
the free part contributing only for $\omega\gtrsim\omega_0$,
as opposed to contributing for $\omega>0$.
The results for a number of cuts with different $\omega_0$ applied in the 
fit procedure is shown in Tab.~\ref{tab_fit_res_cut}. A value of 
$\Delta_{\omega}/T=0.5$ is used throughout the scan; the results are 
insensitive to its actual choice. The results for 
$c_{BW}T/\Gamma \sim \sigma$ are rising slightly when moving the cut to
higher frequencies, showing that the peak rises in height. Around 
\begin{table}[h]
	\centering
	\begin{tabular}{|c|c|ccccc|}
		\hline 
		$\Delta_{\omega}/T$ & $0.0$ & \multicolumn{5}{|c|}{$0.5$} \\ 
		$\omega_0/T$ & $0.0$ & $1.0$ & $2.5$ & $3.0$ & $3.5$ & $4.0$ \\ \hline
		$\Gamma/T$ & $2.89(1.12)$ & $2.85(1.09)$ & $2.99(0.95)$ & $3.31(0.91)$ & $3.88(0.88)$ & $4.75(0.88)$ \\ 
		$\frac{c_{BW}T}{\Gamma}$ & $0.524(146)$ & $0.543(149)$ & $0.607(138)$ & $0.610(115)$ & $0.595(88)$ & $0.571(44)$ \\
		$k$ & $0.039(7)$ & $0.039(7)$ & $0.038(7)$ & $0.038(7)$ & $0.037(7)$ & $0.035(7)$ \\ \hline
		$\chi^2/\text{dof}$ & $1.24$ & $1.24$ & $1.23$ & $1.22$ & $1.21$ & $1.19$ \\ \hline
	\end{tabular}
	\caption{Fit results for $T=1.1T_c$ and some selected cutoffs. 
		Note that $c_{BW}/\Gamma$ is 
	directly proportional to the electrical conductivity $\sigma$.}
	\label{tab_fit_res_cut}
\end{table}
$\omega_0/T \simeq 3$ the peak becomes much broader to compensate for
the cut off contribution and $c_{BW}T/\Gamma$ falls of again. At this point
the Breit-Wigner peak contributes (as the only contribution) 
to a frequency regime that is, for the uncut fit, already dominated
by the free part, see Fig.~\ref{fig_spf_plot}. Raising $\omega_0/T$
even further does, from a physical point of view, not make sense anymore. 
In addition, note how the 
value of $\chi^2/\text{dof}$ is not rising althrough the procedure: a-priori
there is no reason for the fit to become much worse, in terms of its $\chi^2$ value,
upon the application of 
such cuts. It turns out the ansatz can perfectly compensate for the missing free 
contribution to the extent of $\omega_0/T \simeq 3$. Beyond that, mathematically
the Breit-Wigner peak can still compensate for the cut, but, as argued above, 
the initial 
physical motivation of this form of ansatz is not given anymore.
\newline

\begin{figure}[h]
	\centering
	\includegraphics[width=0.495\textwidth]{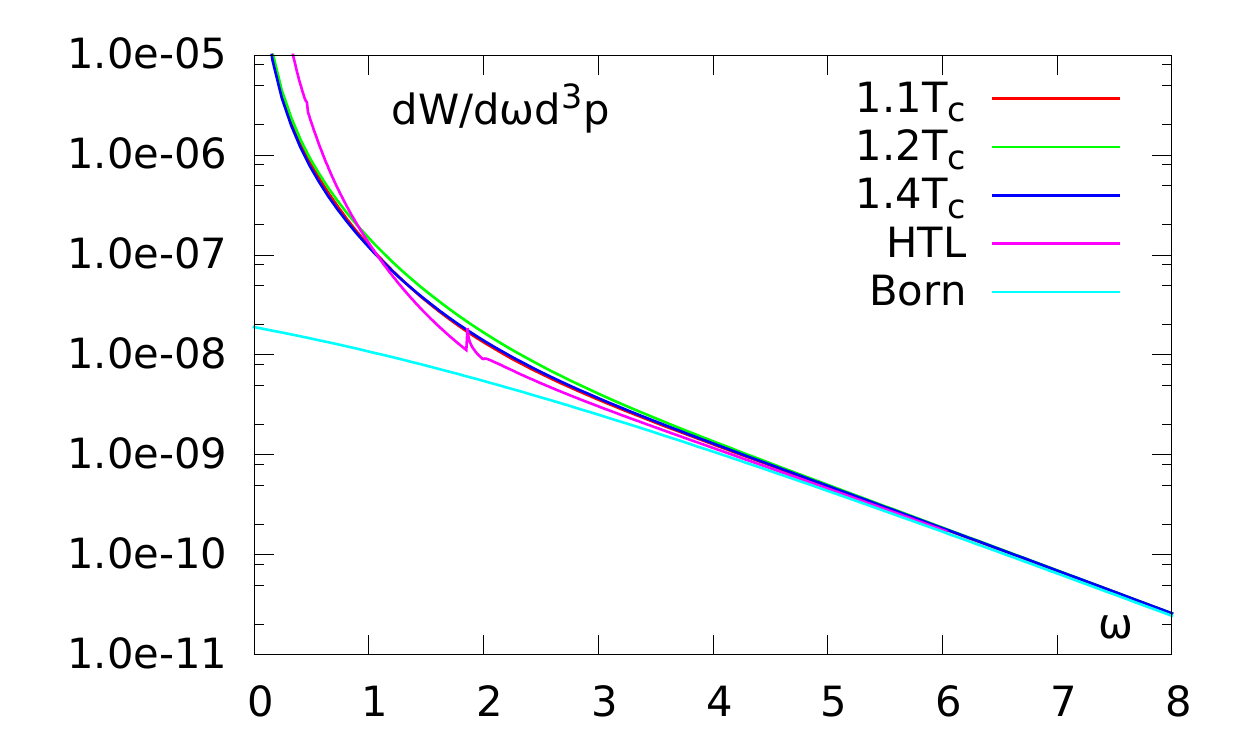}
	\caption{The thermal dilepton rate as a function of $\omega$.}
	\label{fig_dilrate}
\end{figure}

\begin{table}[h]
	\centering
	\begin{tabular}{|c|ccc|}
		\hline
		$T$ & $1.1T_c$ & $1.2T_c$ & $1.4T_c$ \\ \hline
		$\left(\frac{\sigma}{C_{em}T}\right)_{max}$ & $0.436$ & $0.365$ & $0.361$ \\ 
		$\left(\frac{\sigma}{C_{em}T}\right)_{min}$ & $0.217$ & $0.211$ & $0.217$\\ \hline
	\end{tabular}
	\caption{Final results for the electrical conductivity. Note that the 
		systematic error from the cut-procedure and the statistical fit
	error are included (see text).}
	\label{tab_sigma_res}
\end{table}
For the electrical conductivity, however, we can include its maximal
deviation from the uncut result as an upper systematical error. Our results
for the electrical conductivity for all three temperatures are given in
Tab.~\ref{tab_sigma_res} and Fig.~\ref{fig_spf_plot} (\textit{right}), respectively.
They are comparable to recent studies \cite{Brandt2013,Amato2013}
using MEM and Wilson Clover fermions at finite lattice spacing.
For a comparison of different calculations of the electrical conductivity
see \cite{Greiner2010}.
The thermal dilepton rates calculated from our ansatz for the spectral functions
via the first expression of (\ref{eqn_dilrate})
are shown in Fig.~\ref{fig_dilrate}
for all three temperatures.
They are qualitatively comparable to the rate obtained by an HTL calculation
\cite{Braaten1990} in the intermediate $\omega$ region, as well as to the 
leading order (Born) rate for large $\omega$. However, our results show a 
better behaviour for small $\omega$ consistent with a finite electrical 
conductivity (see also Fig.~\ref{fig_spf_plot} (\textit{left})).

\section{Conclusion}
Using non-perturbatively improved Wilson Clover valence fermions we 
performed continuum extrapolations of light vector channel correlation
functions. The extrapolations yield reliable results
with errors at the sub-percent level. 
Employing an ansatz for the corresponding spectral function, these 
are used to perform
a fully correlated $\chi^2$-minimization and to
obtain results for the spectral functions and the electrical 
conductivities via a Kubo relation. The electrical conductivities are in accordance 
with earlier results obtained by MEM and $\chi^2$-minimization methods. 
The thermal dilepton rates are obtained and compared to the HTL and leading
order rates and show almost no thermal effect in the analyzed temperature 
region. \\
{\tt Acknowledgements:} The results have been achieved using the PRACE Research Infrastructure
resource JUGENE based at the J\"ulich Supercomputing Centre in Germany and the
Bielefeld GPU-cluster resources. 
This work has been partly supported by BMBF under grants 05P12PBCTA and
56268409 and the GSI BILAER grant.


\end{document}